# Acquisition and Analysis of Crowd-Sourced Traffic Data

**Markus Hilpert**
Department of Environmental Health Sciences, Mailman School of Public Health, Columbia University

**Jenni A. Shearston**
Department of Environmental Health Sciences, Mailman School of Public Health, Columbia University

**Jemaleddin Cole**
Website Developer

**Steven N. Chillrud**
Lamont-Doherty Earth Observatory of Columbia University

**Micaela E. Martinez**
Department of Environmental Health Sciences, Mailman School of Public Health, Columbia University

***Abstract*** Crowd-sourced traffic data offer great promise in environmental modeling. However, archives of such traffic data are typically not made available for research; instead, the data must be acquired in real time. The objective of this paper is to present methods we developed for acquiring and analyzing time series of real-time crowd-sourced traffic data. We present scripts, which can be run in Unix/Linux like computational environments, to automatically download tiles of crowd-sourced Google traffic congestion maps for a user-specifiable region of interest. Broad and international applicability of our method is demonstrated for Manhattan in New York City and Mexico City. We also demonstrate that Google traffic data can be used to quantify decreases in traffic congestion due to social distancing policies implemented to curb the COVID-19 pandemic in the South Bronx in New York City.

## INTRODUCTION

Characterizing road traffic is important because traffic contributes to air pollution and environmental noise and is a measure for human mobility. However, physical measurements of traffic, e.g., using pneumatic tube counters or radar devices, are expensive, particularly at the city scale. Crowd-sourced traffic congestion maps, which are based on the GPS coordinates of cell phones located in driven vehicles, could fill this gap, because the number of users uploading such data to traffic map providers can be large, especially in metropolitan areas, resulting in high spatio-temporal characterization of traffic conditions. Use of crowd-sourced traffic data in environmental research has been pioneered by Hilpert et al. (1) who found that congestion colors displayed on traffic congestion maps are associated with vehicle speed and that traffic congestion is associated with levels of atmospheric black carbon, a tracer for vehicle emissions. Recently Zalakeviciute et al. (2) examined associations between Google traffic data and air quality measurements made with mobile sensors.

While traffic congestion maps are readily displayed on screens of smart phones and computers for real-time illustration of traffic conditions, traffic map archives are typically not accessible to researchers. To use the data for research purposes, traffic maps must be saved or evaluated in real time, while they are displayed on digital devices. However, it is not straightforward to save the screen content to an image file for later analysis, because maps which include a traffic layer are accessed and displayed on personal digital devices through external JavaScripts$^{TM}$ that reside on servers of the traffic data providers (3). These scripts cannot be modified by a user to save the map in an image file.

The objective of this paper is to detail a method and scripts for acquiring real-time crowd-sourced traffic congestion data for a latitude/longitude defined map area of interest composed of an array of adjacent tiles. We demonstrate international applicability of our approach by obtaining traffic maps from Google, a popular provider of crowd-sourced traffic maps, for the Metropolitan areas of Mexico City and New York City. We also demonstrate one application of these data, showing that time series of traffic maps reflect the expected changes in traffic due to social distancing policies implemented to curb the COVID-19 pandemic in New York City.

**DATA AQUISTION**

**Overview**

We developed and present here detailed instructions for downloading an array of Google maps with a traffic layer which we loosely call traffic congestion maps. This task can be split up into five steps:

- Definition of traffic map area: Define the geometry of the rectangular array of square tiles for which traffic data is requested.

- Displaying a Google traffic map: Modify the HTML script we developed for displaying one Google traffic map tile to account for the Application Programming Interface (API) key of the user.

- Downloading a Google traffic map array: Adjust the C shell script we

developed for downloading a Google traffic map array to account for the map area defined in Step 1 and the directory structure on a user's computer with a Unix/Linux like computing environment.

- Automating data download of time series: Automate a call of the C shell script at predetermined points in time to obtain time series of the traffic maps.

- Merging downloaded traffic tiles: Stitch the final array of image files together to form one image for the entire study area for subsequent analysis.

Each of these steps is described in detail in the following subsections.

**Definition of traffic map area**

To account for the typically non-square shape of a region of interest, we define the area for which traffic data is requested by a rectangular array of square traffic map tiles. Our scripts require specifying the following quantities: 1) the coordinates of the center of the map array, $(\lambda_c, \phi_c)$, where $\lambda_c$ is the latitude angle and $\phi_c$ is the longitude angle, 2) the number of pixels $N_{pix}$ along one side of the square tile, 3) the integer zoom level $z$ of the square tiles, and 4) the number of tiles along the latitude and longitude lines, $N_{lat}$ and $N_{long}$.

The zoom level $z$ determines the size of the square tiles. To understand this dependence, one needs to understand first the spatial resolution of Google maps. This topic has been explained well in an exchange on a GIS-related question posted on StackExchange (4) and is summarized next. Google maps divide the surface of the Earth into square pixels ($\Delta x$) delineated by lines of latitude (angle $\lambda$) and longitude (angle $\phi$). Google discretizes the equator ($\lambda = 0°$ latitude line) by $2^{32}$ = 4,294,967,296 pixels at a zoom level $z = 24$. From the equatorial radius of the Earth, $R_{Earth} = 6,378,137$ m (5), one can calculate the pixel size along the equator at that zoom level: $\Delta x_{24} = 2\pi R_{Earth} / 2^{32} = 9.3$ mm. For other zoom levels z, the pixel size along the equator is given by

$$\Delta x_z = \Delta x_{24} \, 2^{24-z} \tag{1}$$

Hence, decreasing $z$ by one doubles the pixel size. Since the number of pixels along all lines of latitude must be the same for a given zoom level, the pixel size at any latitude angle $\lambda$ depends on $\lambda$ and is given by

$$\Delta x_z(\lambda) = \Delta x_{24} \, 2^{24-z} \cos \lambda \tag{2}$$

From the pixel size one can now calculate the side length of each square tile:

$$l = N_{pix}\, \Delta x_z(\lambda) \tag{3}$$

In our applications we found $N_{pix} = 1{,}000$ to be a practical choice. For instance, at a zoom level $z = 15$ at which traffic on the smallest streets is resolved in New York City ($\lambda = 40.7°$), the size of the map tile is $l = 3.6$ km. The script, however, can be modified according to a user's specific geographic location and needs.

To exclude certain traffic tiles from download, it is useful to define the latitude and longitude angles $\lambda_i$ and $\phi_j$ of all array tiles:

$$(\lambda_i, \phi_j) = (\lambda_c, \phi_c) + \left((i - (N_{lat}+1)/2)\,\Delta\lambda,\ (j - (N_{long}+1)/2)\,\Delta\phi\right) \tag{4}$$

where the increments $\Delta\lambda$ and $\Delta\phi$ determine the size of the tiles, and the indices $i = 1, \ldots, N_{lat}$ and $j = 1, \ldots, N_{long}$ specify the map tiles within the rectangular map array. In our scripts, these indices are used to exclude traffic tiles. For example, the index pair $(i=1, j=1)$ denotes the tile in the Southwest corner of the array.

For the sake of clarity, we also comment on the calculation of $\Delta\lambda$ and $\Delta\phi$, which do not need to be specified by users of our scripts but are calculated internally within the scripts. Since $l$ can be interpreted as an arc length along the circle defined by the $\lambda$ latitude line, the increment in latitude angle can be calculated as follows:

$$\Delta\lambda = \frac{l}{2\pi\, R_{Earth}\, \cos\lambda}\, 360° \tag{5}$$

Similarly, since $l$ can be interpreted as an arc length along the circle defined by the $\phi$ longitude line, the increment in longitude angle is given by

$$\Delta\phi = \frac{l}{2\pi\, R_{Earth}}\, 360° \tag{6}$$

**Displaying a Google traffic map**

While it is our objective to save a Google traffic map as an image file for subsequent analysis, displaying such a traffic map is a first step toward achieving this goal and also helps with debugging the scripts and defining the mapped area. We adapted an existing HTML script published on the Google website for displaying a traffic map for arbitrary coordinates and zoom level (3). Latitude and longitude of the map center as well as the zoom level are passed to our HTML script entitled "Load_Traffic_Map.html" through so-called URL parameters. **Figure A** in the Appendix shows our HTML script. It can be invoked (and debugged) by copying the following URL into the URL address bar of a web browser:

file:////work_directory/Load_Traffic_Map.html?lat=33.73&long=-84.43$z=15

where the directory path needs to be modified to account for the directory structure of a user's computer account. The latitude and longitude angles (termed "lat" and "long" in the URL) as well as the zoom level (termed "z") can be adjusted too. Note that a user must obtain an application programming interface (API) key from Google in order to obtain the requested map, and then replace "USERKEY" by that API key in the file "Load_Traffic_Map.html". The API key is also used to charge the credit card of the user for the requested data.

**Downloading traffic map array**

Many traffic map providers, including Google, provide tools for displaying traffic maps on mobile devices (e.g., smart phones) and desktop computers. This is accomplished via an external JavaScript located on a Google server that processes input from the user's device (e.g., to move the map or to change the zoom level). However, this JavaScript does not allow a user to save the map to an image file. To address this issue, we use a headless internet browser, a command line tool which website developers typically use for the purposes of website testing. Some headless browsers allow saving the website that would be displayed if a regular browser was used to an image file. Not all headless internet browsers are compatible with Google map websites. One browser that works for our application is the Google Chrome$^{TM}$ browser, which can be run both with a Graphical User Interface (GUI) or in headless mode.

**Figure B** in the Appendix shows the C shell script entitled "Download_Traffic_Map_Array" we developed to download an array of traffic maps and save them to image files (png format). The output image file name contains information about date and time of the download and the position of each map tile within the array based on the indices $i$ and $j$ in Equation (4). The script expects the user to provide the following information: the output directory for the traffic image files; the location of the "Google Chrome" internet browser including complete directory path information; the location of the HTML script "Load_Traffic_Map.html", including the complete directory path information; the coordinates of the center of the map array $(\lambda_c, \phi_c)$; the number of tiles, $N_{lat}$ and $N_{long}$, along the vertical and horizontal direction of the map array; optional specification of N tiles that should be excluded from the download by the two N-dimensional vectors of integers, which contain the integer indices $i$ and $j$ of the tile coordinates according to Equation (4); the zoom level $z$; and the number of pixels along one side of a square tile, $N_{pix}$. The remainder of the script can be modified by users according to specific needs and preferences.

While obvious to experienced C shell users, we comment on some elements of the script: 1) Since the C shell script only allows performing calculations involving integer variables, we used the unix command line tool "awk" to perform floating-point calculations according to Equations (1) through (6); and 2) To pass the question mark "?", which separates the URL from the URL parameters to the Google traffic server, we set the "noglob" flag because "?" is otherwise interpreted as a wildcard.

**Automating data download of times series**

We used the Unix "cron" daemon to execute the "Download_Traffic_Map_Array" C shell script periodically. The "cron" daemon is a process that runs continuously in the background and allows scheduling execution of commands at specific dates and times. "cron" can be configured by the "crontab" command. For instance, to execute our script every hour, a user can call the command "crontab -e" from the Unix command prompt, and type or paste the following line into the "vi" editor (6) window that opens upon execution of the command:

0 * * * * work_directory/Download_Traffic_Map_Array

The cron daemon then executes the C shell script periodically (every hour in this example) for an indefinite period of time, until the user deletes the line above by again invoking the "crontab -e" command. The flow chart shown in **Figure 1** summarizes the steps that need to be taken to collect traffic data.

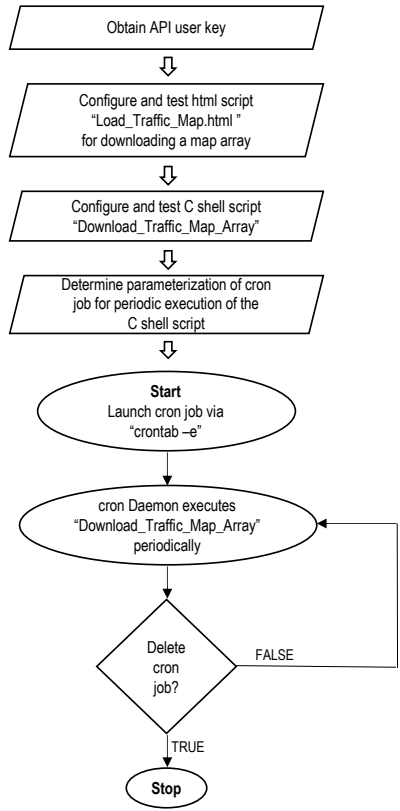

**Figure 1.** Flow chart for traffic data collection. Wide arrows indicate steps that need to be taken to prepare program execution. Lean arrows indicate actual program execution.

**Merging downloaded traffic tiles**

We used the commercial Matlab$^{TM}$ software to stitch together the individual traffic map tiles and to display them on a computer monitor. To that end, we developed the Matlab script entitled "Read_Traffic_Map_Array.m" (see **Figure C** in the Appendix). It can be invoked from the Matlab command line by typing the following command:

D=Read_Traffic_Map_Array('06_01_20','09:00');
imshow(D)

The first and second parameter of the function call, quoted in the parentheses, denote date and time when the script "Download_Traffic_Map_Array" was invoked. Note that the Matlab work directory needs to be set to the directory in which the image files are present.

**Figure D** in the Appendix shows a comparable script we wrote in the free statistical "R" software (version 3.5.0). We note that the Matlab script is executed substantially faster than the R script.

## SAMPLE APPLICATIONS

### International applicability

We provide two examples demonstrating the broad and international applicability of our method. **Figure 2** shows a traffic congestion map obtained for Manhattan in New York City. Twelve map tiles were obtained at a zoom level $z = 15$. At that zoom level, traffic on the smallest streets is resolved in this high-resolution map when zooming in as illustrated by the magnified circular region. In this example, we excluded six tiles not containing the island of Manhattan from the download.

**Figure 3** shows a traffic congestion map for Mexico City. Nine map tiles were obtained at a zoom level $z = 13$. At that zoom level, only traffic on main arteries is provided by the Google traffic server, and thus data for the smallest streets is not represented in the downloaded maps.

### Changes in traffic due to the COVID-19

To demonstrate that time series of traffic maps can be used to detect and characterize changes in traffic, we analyzed changes in traffic in New York City, one of the early epicenters of the COVID-19 pandemic in the US, following implementation of social-distancing policies to curb the pandemic. We focused on the South Bronx, where we previously examined associations between real-time measurements of airborne black carbon and congestion colors displayed on crowd-sourced traffic maps. These analyses were performed at three Sites: Site A, a one-way off-ramp of an interstate; Site B, a small one-way street in a mixed-use area; and Site C: a two-way street which provides relatively direct access to the entrance to a distribution hub of a shipping company and an online grocery store warehouse. The three sites are highlighted in the magnified region in Figure 2.

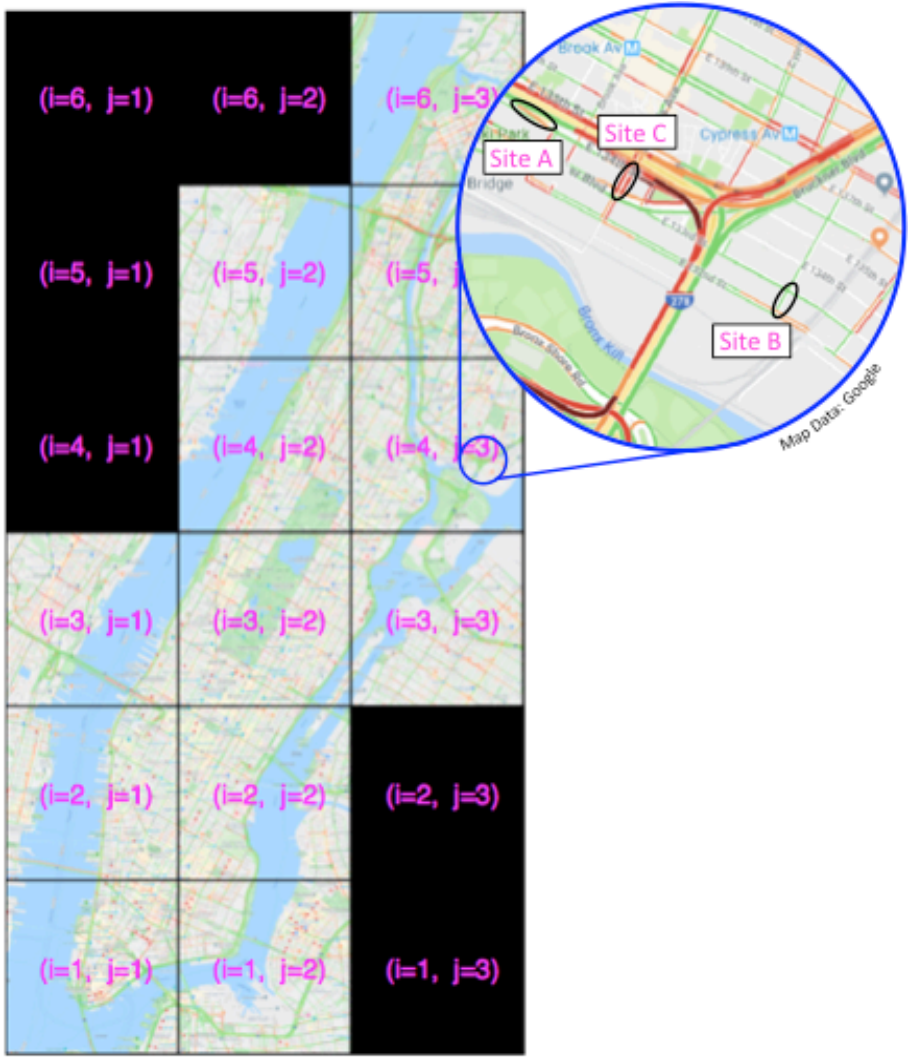

**Figure 2.** Sample traffic map for a portion of New York City including Manhattan and the South Bronx at zoom level *z* = 15. The black ovals in the magnified region delineate four road segments at Sites A, B and C for which we analyzed time series of traffic congestion colors. Black raster lines are not part of the actual map array and were added to delineate map tiles. The magenta text is also not part of the map and has been added to illustrate the indices used to specify map tiles according to Equation (4).

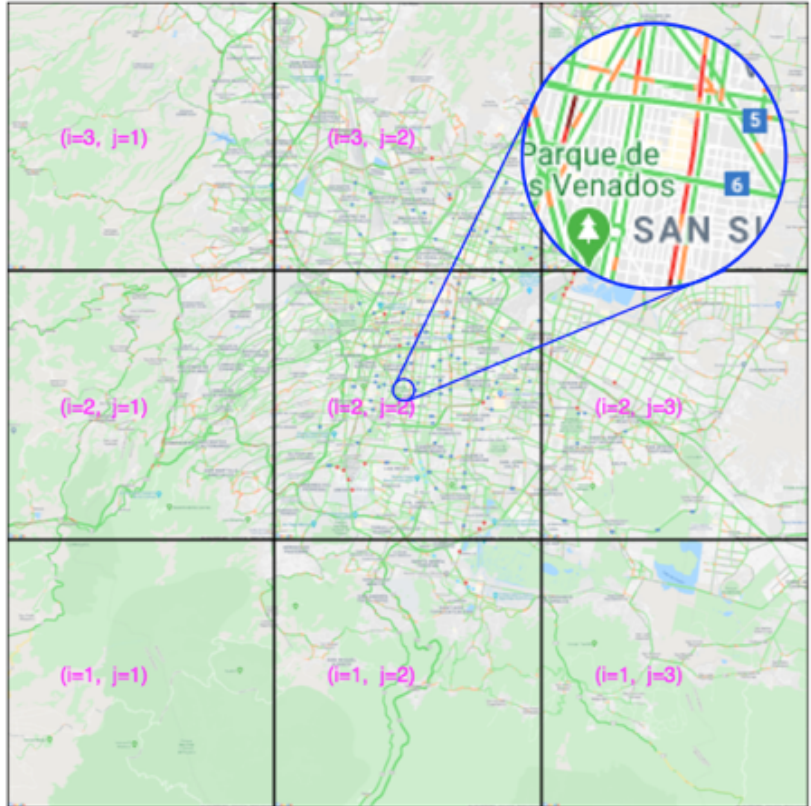

**Figure 3.** Sample traffic map for Mexico City at *z* = 13. Black raster lines are not part of the actual map array and were added to delineate the individual map tiles. The magenta text is also not part of the map and has been added to illustrate the indices used to specify map tiles according to Equation (4). Map data: Google.

We acquired traffic maps during New York City's stay-at-home-order, which lasted about 11 weeks from March 22 to June 22, 2020, to be then subsequently relaxed in four reopening phases. To evaluate changes in traffic conditions, we also analyzed baseline traffic maps acquired before the stay-at-home-order for a ~12-week period lasting from January 1 to March 22, 2020. We acquired images eight times a day, every three hours, and determined the congestion colors shown on the traffic maps for the three sites described above. For the three sites, we determined the time series of the ordinal congestion color code (1):

$$\text{CCC} = \begin{cases} \boxed{1} \\ \boxed{2} \\ \boxed{3} \\ \boxed{4} \\ \boxed{5} \end{cases}$$

where the color of the square box indicates the congestion color displayed on traffic maps (maroon, red, orange, green and gray), and the number in the box indicates CCC. Decreasing CCC values (increase in redness) indicate increased congestion.

For the one-way road segments of Sites A and B, one CCC time series was obtained for each site. For the two-way road segment of Site C, two CCC time series were obtained, because on the traffic maps two parallel colored road segments are shown to describe traffic in the Northbound and Southbound directions. Thus, four CCC time series for four road segments in the traffic map were obtained and analyzed in total.

**Figure E** in the Appendix visualizes the times series of the 12-hour CCC (averaged over 4 consecutively taken maps) for the four road segments. (We do not show the 3-hour time series because then CCC would have assumed integer values, causing jagged curves which hinder the visualization of temporal trends.) At all sites, CCC increased, both on weekdays and weekends, following the stay-at-home order, indicating overall less traffic congestion at the sites (**Table 1**). Specifically on weekdays, the increase in CCC averaged across the four road segments was 0.19, and these increases were significant for all road segments, except for Southbound traffic at Site C. On weekends, the CCC increase averaged across road segments was 0.13, and increases were significant at only half of the road segments.

**Table 1.** Changes in congestion color code CCC (mean and standard deviation) at four road segments in the South Bronx before and after implementation of a stay-at-home order. P-value indicates the significance of the difference in CCC between the baseline and stay-at-home periods.

|  | Weekdays | | | Weekends | | |
| --- | --- | --- | --- | --- | --- | --- |
|  | CCC (Baseline) | CCC (Stay-at-home) | p | CCC (Baseline) | CCC (Stay-at-home) | p |
| **Site A** | 3.61 (0.22) | 3.81 (0.16) | 6e-6 | 3.87 (0.15) | 3.90 (0.13) | .6 |
| **Site B** | 3.88 (0.25) | 4.16 (0.26) | 5e-6 | 3.78 (0.30) | 4.10 (0.29) | .003 |
| **Site C** | | | | | | |
| Southbound | 2.99 (0.21) | 3.05 (0.28) | 0.24 | 3.02 (0.25) | 3.21 (0.31) | 0.05 |
| Northbound | 3.09 (0.25) | 3.20 (0.25) | 0.04 | 3.14 (0.29) | 3.20 (0.25) | 0.5 |

**DISCUSSION**

We have presented in detail a method and scripts for acquiring arrays of Google traffic congestion maps. Our method can be applied anywhere in the world where Google traffic maps are available, and is therefore widely applicable. In regions where no Google maps are available (e.g., China), our method can potentially be modified to acquire data from other traffic data providers (e.g. Apple Maps).

Our scripts have been tested on three different Apple computers running the OS X operating systems 10.11.6 or 10.12.6. These scripts can be expected to run in other Linux/Unix computational environments, because we used command line tools that are native to these operating systems. However, slight modifications might be necessary, particularly when it comes to directory structure.

For Google maps, the road segments in the downloaded images can be segmented into five congestion colors: gray, green, orange, red, and maroon as previously demonstrated (1), where the order indicates increasing degree of congestion, which is related to the capacity of the particular road type. Time series of the congestion color code CCC of the road segments, an ordinal representation of the five congestion colors, can be used in environmental research.

We used traffic congestion data, which we acquired with the methods presented in this paper, to characterize the effects of a public health intervention on road traffic. We specifically examined changes in traffic at three road sites in the South Bronx in response to New York City's stay-at-home order implemented to curb the COVID-19 pandemic. Traffic congestion was quantified in terms of the congestion color code CCC. Overall, the increases in CCC and hence decreases in congestion due to the stay-at-home order were both larger and more significant on weekdays compared to weekends across the three sites, perhaps because the stay-at-home order reduced human mobility more on weekdays than on weekends, e.g., due to school closures. The fact that p-values for the CCC increases at Site C tend to be much larger than for Sites A and B could potentially be attributed to the fact that Site C is relatively close to essential businesses which might have even seen increased activity during the pandemic, such as the nearby shipping hub and the online grocery store (7). The decrease in traffic congestion due to the COVID-19 pandemic we observed in the South Bronx is consistent with COVID-19-related decreases in traffic in New York City (8).

## CONCLUSION

Crowd-sourced traffic data represent a novel, widely available, and highly spatio-temporally resolved data source for environmental research. We presented a methodology and scripts for obtaining such data.

## ACKNOWLEDGMENT

MEM and MH were supported by the NSF Award Number 2029421, RAPID: Transmission and Immunology of COVID-19 in the Pandemic and Post-Pandemic Phase: Real-time Assessment of Social Distancing & Protective Immunity. MH and SNC were also supported in part by NIEHS grants R21ES030093 and P30ES009089. MEM was also supported by the Office of the Director, National Institutes of Health Health under award number DP5OD023100. JAS was supported by NIEHS grant T32 ES007322.

# APPENDIX

```html
<!DOCTYPE html>
<html>
<head>
  <meta name="viewport" content="initial-scale=1.0, user-scalable=no">
  <meta charset="utf-8">
<title>Traffic Layer</title>
<style type="text/css">
  #map { height: 100%;         }
  html, body { margin: 0;
             padding: 0;       }
</style>
<script type="text/javascript">
const HeadParam = new URLSearchParams(document.location.search);
const n = HeadParam.get("n");
console.log(n);
var stylesheet = document.styleSheets[0];
stylesheet.cssRules[0].style.height=n+"px";
stylesheet.cssRules[0].style.width=n+"px";
</script>
</head>

<body>
<div id="map"></div>
<script>
function initMap()
{ const params = new URLSearchParams(document.location.search);
  const Lat = params.get("lat");
  const Long = params.get("long");
  const Zoom = params.get("z");
  var map = new google.maps.Map(document.getElementById('map'),
      { zoom: parseFloat(Zoom),
        center: {lat: parseFloat(Lat), lng: parseFloat(Long)},
        disableDefaultUI: true
      });
  var trafficLayer = new google.maps.TrafficLayer();
  trafficLayer.setMap(map);
}
</script>

<script src="https://maps.googleapis.com/maps/api/js?key=API_userkey&callback=initMap">
</script>

</body>
</html>
```

**Figure A.** HTML code/file "Load_Traffic_Map.html" for displaying a Google traffic map for a user-specifiable LatLong coordinates. Note that USERKEY must be replaced by a valid API key. The field to be edited by a user is highlighted in yellow.

```csh
#!/bin/csh

# user should modify the following parameters
# directory to which image files should be saved:
cd /Users/user_name/work_directory/IMAGES
# specify location of "Google Chrome" internet browser
alias chrome '/Applications/Google\ Chrome.app/Contents/MacOS/Google\ Chrome'
# specify complete path to HTML script
set Script = \
    'file:////Users/user_name/HTML_Scripts/Load_Traffic_Map.html'
# LatLong of the center of map array (for Manhattan, NYC):
set Lat_c  = 40.79
set Long_c = -73.97
# number of latitude and longitude tiles of downloaded map array:
set N_lat = 6
set N_long = 3
# Excluded tiles:
set X_long = (1 1 1 2 3 3)
set X_lat  = (4 5 6 6 1 2)
# Zoom level:
set Z = 15
# Number of pixels along a side of a square traffic map tile
set N_pix = 1000

# the following code can be modified if needed
set R_Earth = 6378137
set dx_Equator = 0.009330692
set COS = `echo $Lat_c | awk '{printf "%.15f\n",cos($1*atan2(0,-1)/180)}'`
set Circumference_Equator = \
    `echo $R_Earth | awk '{printf "%.15f\n",2*atan2(0,-1)*$1}'`
set dx_lat = \
    `echo $Z $dx_Equator $COS | awk '{printf "%.15f\n", $3 * $2 * 2^(24-$1)}'`
set d_theta=`echo $N_pix $dx_lat $Circumference_Equator | \
    awk '{printf "%.15f\n",$1*$2/$3*360}'`
set d_phi=`echo $d_theta $COS | awk '{printf "%.15f\n",$1/$2}'`
set TimeStamp="`date +%m_%d_%y__%H:%M`"
set noglob
foreach i_lat (`seq 1 1 $N_lat`)
 foreach i_long (`seq 1 1 $N_long`)
   set Download=1
   if ($#X_long>0) then
     foreach m (`seq 1 1 $#X_long`)
       if (($i_lat == $X_lat[$m]) && ($i_long == $X_long[$m])) then
          set Download=0
       endif
     end
   endif
   if ($Download == 1) then
     set Long = `echo $Long_c $i_long $d_phi $N_long | \
                 awk '{printf "%.10f\n", $1+(-0.5*$4-0.5+$2)*$3}'`
     set Lat = `echo $Lat_c $i_lat $d_theta $N_lat | \
                 awk '{printf "%.10f\n", $1+(-0.5*$4-0.5+$2)*$3}'`
     set URL=$Script'?lat='$Lat'&long='$Long'&z='$Z'&n='$N_pix
     echo $URL
     set OutFile = "TrafficMap_"$i_lat"_"$i_long"_"$TimeStamp".png"
     chrome --headless --virtual-time-budget=10000000 \
            --disable-gpu --window-size=$N_pix,$N_pix \
            --screenshot=$OutFile $URL
   endif
 end
end
```

**Figure B.** C shell script "Download_Traffic_Map_Array" for obtaining an array of traffic maps using the html script from Figure 1 and saving them to image files. Fields to be edited by a user are highlighted in yellow.

```
################################
#### Read Traffic Map Array ####
################################
# First, set working directory to location where image files are saved.
# setwd("")
# Second, ensure the image files are in one folder.
# Third, install the R package "imager" and load it
# install.packages("imager")
library(imager)

Read_Traffic_Map_Array = function(Date, Time){
  
  if (missing(Date)) {
    stop('Usage: Read_Traffic_Map_Array("09_19_19", "10_30")')
  } else if (missing(Time)) {
    stop('Usage: Read_Traffic_Map_Array("09_19_19", "10_30")')
  } else if (!is.character(Date)) {
    stop('Usage: Read_Traffic_Map_Array("09_19_19", "10_30")')
  } else if (!is.character(Time)) {
    stop('Usage: Read_Traffic_Map_Array("09_19_19", "10_30")')
  }
  NXY = 2000
  Files = list.files(getwd(),
              pattern=paste0("TrafficMap_._._",Date,"__",Time,".png"))
  N_lat  = 0
  N_long = 0
  for (idx in 1:length(Files)) {
    Files[idx]
    Lat = substr(Files[idx],12,12)
    Long = substr(Files[idx],14,14)
    N_lat = max(N_lat, as.numeric(Lat))
    N_long = max(N_long, as.numeric(Long))
  }
  Merge = array(data = 0, dim = c(N_long*NXY, N_lat*NXY, 1, 4))
  Merge = as.cimg(Merge)
  for (idx in 1:length(Files)) {
    D = load.image(Files[idx])
    Lat = as.numeric(substr(Files[idx],12,12))
    Lat = N_lat - (Lat-1)
    Long = as.numeric(substr(Files[idx],14,14))
    Merge[(1+(Long-1)*NXY):(Long*NXY),(1+(Lat-1)*NXY):(Lat*NXY), , ] = D
  }
  plot(Merge)
  save.image(Merge,"TrafficMapArray.png")
}
```

**Figure C.** Matlab script for displaying a downloaded array of Google traffic maps.

```
function Merge=Load_Traffic_Map_Array(Date,Time);

if nargin==0
  disp('Usage:  Load_Traffic_Map_Array(''09_19_19'',''10:30'')')
  return
end
St = ['TrafficMap*',Date,'__',Time,'.png'];
Files = dir(St);
N_lat  = 0;
N_long = 0;
for idx=1:size(Files,1)
  disp(Files(idx).name);
  Lat =Files(idx).name(12);
  Long=Files(idx).name(14);
  N_lat =max(N_lat,str2num(Lat));
  N_long=max(N_long,str2num(Long));
  if idx==1
    D=imread(Files(idx).name);
    N_pix=size(D,1);
    disp(['N_pix = ',num2str(N_pix)])
  end
end
Merge = uint8(zeros(N_lat*N_pix,N_long*N_pix,3));
for idx=1:size(Files,1)
  D=imread(Files(idx).name);
  Lat =str2num(Files(idx).name(12));
  Long=str2num(Files(idx).name(14));
  Merge(1+(N_lat-Lat)*N_pix:(N_lat-Lat+1)*N_pix,1+(Long-
1)*N_pix:(Long)*N_pix,:) = D;
end
```

**Figure D.** R script for displaying a downloaded array of Google traffic maps.

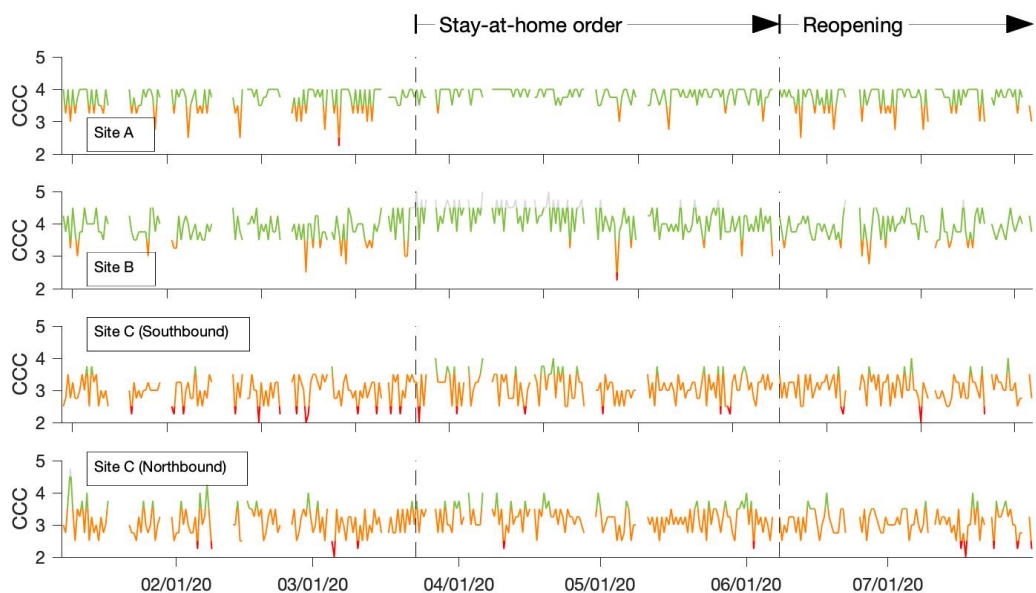

**Figure E.** Time series of congestion color code CCC for three road segments in the South Bronx (see Figure 2): 1) an interstate off-ramp (Site A), 2) a small one-way street (Site B), and 3) a two-way road (Site C). The colors of the lines indicate the color of the road segments displayed in the traffic congestion map. Lower values represent decreased speed and increased congestion.